# Controlled coalescence-induced droplet jumping on flexible superhydrophobic substrates

Gopal Chandra Pal[†], Siddharth SS[†], Manish Agarwal[‡], Chander Shekhar Sharma[†]*

[†]Thermofluidics Research Lab, [‡]Department of Mechanical Engineering, Indian Institute of Technology Ropar, Rupnagar, Punjab 140 001, India
*Email: chander.sharma@iitrpr.ac.in, Ph: +91-1881-232358

## Abstract

Sessile droplets coalescing on superhydrophobic substrates spontaneously jump from the surface. In this process, the excess surface energy available at the initiation of coalescence overcomes the minimal surface adhesion and manifests as sufficient kinetic energy to propel the droplets away from the substrate. Here, we show that the coalescence induced droplet jumping velocity is significantly curtailed if the superhydrophobic substrate is flexible in nature. Through detailed experimental measurements and numerical simulations, we demonstrate that the droplet jumping velocity and jumping height can be reduced by as much as 40 % and 64%, respectively, by synergistically tuning the substrate stiffness and substrate frequency. We show that this hitherto unexplored aspect of droplet coalescence jumping can be gainfully exploited in water harvesting from dew and fog harvesting. Additionally, through an exemplar butterfly wing substrate, we demonstrate that this effect is likely to manifest on many natural superhydrophobic substrates due to their inherent flexibility.

## Introduction

Coalescence induced droplet jumping manifests for small size droplets on superhydrophobic substrates. In this coalescence process, the excess surface energy overcomes the minimal contact line dissipation and causes the coalesced droplets to jump off the surface(*1*). This passive, gravity independent, droplet removal from the surface is critical to a plethora of applications, such as increasing the condensation heat transfer efficiency of the industrial condenser (*2*, *3*), phase change cooling for thermal management of electronics (*4*, *5*), creating anti-icing surface (*6*), energy harvesting (*2*, *7*) and water harvesting (*8–10*). Hence, many experimental(*1*, *3*, *11–20*) and numerical (*11*, *13*, *15–17*, *21–29*) studies have been carried out to investigate the effects of parameters, such as surface macrotextures and wettability, on the overall efficiency of the jumping process(*11*, *12*).

Here, we investigate coalescence induced droplet jumping on flexible nanotextured superhydrophobic substrates. Coalescing droplets exert a transient force on the substrate (*19*), and we demonstrate that this force, and consequently the jumping velocity can be controlled by tuning the substrate stiffness and natural frequency. We analyze individual droplet coalescence events experimentally and find that droplet jumping velocity can be reduced by as much as 40% for microliter droplets by tuning the substrate stiffness ($k$) and natural frequency ($w_n$). This translates to a nearly 64% reduction in droplet jumping height. This is especially relevant for dew and fog harvesting applications wherein superhydrophobic substrates can achieve high rate of surface renewal due to coalescence induced droplet jumping (*30–32*), but such jumping droplets are vulnerable to loss through entrainment by wind in the vicinity of the substrate (*8*). Thus we experimentally evaluate the effect of substrate flexibility on fog harvesting and find that the use of a flexible substrate can reduce the scattering of jumped droplets. Additionally, we numerically analyze the phenomena using a fluid-structure interaction (VOF-spring mass model) modeling framework that successfully captures the coupling between the droplet coalescence process and the substrate deformation. We also propose a simplified semi-empirical analytical model of the phenomena that helps to elucidate the design of flexible substrate for achieving even higher and on demand reduction in droplet coalescence induced jumping velocity and height. Further, the reduction in droplet jumping velocity is expected to manifest on many natural superhydrophobic surfaces also due to their inherent flexibility. We show this jumping velocity reduction on an exemplar natural surface of a butterfly wing.

**<u>Result and discussion:</u>**

We have prepared flexible super-hydrophobic nanotextured surfaces by spray coating commercially available silanized silica nanoparticles dispersed in isopropanol (Soft 99, Glaco)(*19*, *33*) on thin, flexible sheets of Polydimethylsiloxane (PDMS), copper, and aluminum. Glass slide coated with the same coating has been used as the rigid substrate for comparison. We took care to achieve consistent wettability in terms of advancing contact angle and contact angle hysteresis across the various substrates to isolate the effect of substrate stiffness on droplet coalescence induced jumping. (Refer Materials and Methods). Coalescence-induced droplet jumping experiments are performed by dispensing millimetric droplets onto the substrate using super-hydrophobic micropipette tips (*34*). In each experiment, the substrate is mounted in a fixed-fixed beam configuration (see Figure 2B), and the coalescence is triggered by moving one droplet gently toward the other using a super-hydrophobic tip. The coalescence process is recorded using a high-speed camera, and the captured image sequence is analyzed for droplet motion and substrate deformation. (Refer Materials and Methods and Figure S1).

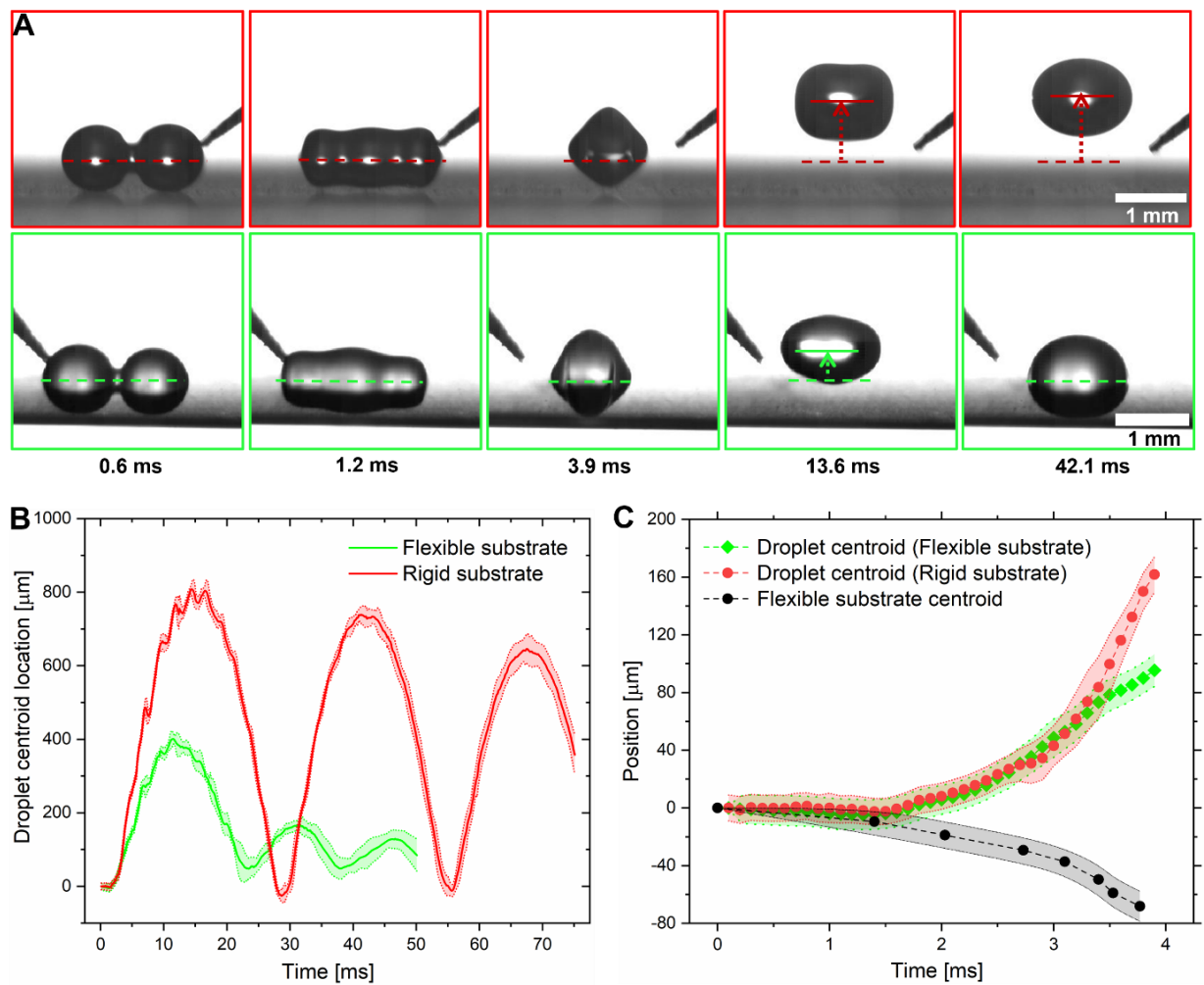

*Figure 1 A: Image sequences comparing coalescence between two droplets of ~1 mm in diameter on horizontal, super-hydrophobic rigid (red) and flexible (green) substrates. The dashed and solid horizontal lines mark the reference and current droplet centroid location, respectively. The reference centroid location corresponds to the initiation of the coalescence process. The flexible substrate is a PDMS beam with thickness, length $L_0$ and width as ~ 52 μm, 15 mm, and 4.2 mm, respectively. B: Temporal evolution of the jumping droplet location for rigid (red) and flexible (green) superhydrophobic substrates corresponding to Figure A. The first peak represents the maximum height attained due to coalescence-induced jumping of the droplet. Subsequent peaks represent the bouncing of the droplet as it returns to the substrate under gravity. C: Magnified view of B, from initiation of coalescence to the moment of droplet jump. Deformation of the flexible substrate (grey) beneath the coalescing droplets is also shown. The three centroid positions are measured relative to their respective initial positions. The shaded area around the curves in B and C represents the uncertainty in the centroid position (refer Supplementary Information Section S2 for details on error analysis).*

In Figure 1, we compare coalescence induced droplet jumping on a flexible superhydrophobic substrate (PDMS) (in green) with that on a rigid surface (in red). The image sequences in Figure 1A show how the two droplet coalescence processes initially evolve almost identically, wherein the neck between the merging droplets expands and comes in contact with the substrate at ~1.2 ms. Subsequently, as the neck continues to expand, the flexible substrate deforms under the force exerted by the coalescing droplets and eventually, the coalesced droplet jumps from the substrate. However, the maximum height reached by the droplet jumping from the flexible surface is nearly 50% lower than that on the rigid substrate, as shown in images at t=13.6 ms. This is highlighted in Figure 1B, which shows the temporal evolution of droplet position subsequent to coalescence induced jumping from the rigid and flexible substrate. (also refer to Supporting Video 1). After reaching maximum height from the rigid substrate, the jumped droplet returns to the surface under gravity, and a series of droplet impact and bounce events follow. It is evident that, in contrast to the rigid substrate case (red curve), the coalesced droplet jumping event and any subsequent droplet bounce on the flexible substrate is significantly suppressed (green curve) due to the deformation of the substrate. Figure 1C illustrates the flexible substrate deformation and droplet movement starting from the initiation of coalescence till the moment of droplet jump from the surface. From the figure, it can be seen there is a negligible movement of the substrate till ~1.2 ms, which is the moment of neck impact on the substrate (refer image sequence in Figure 1A). Beyond this point, the centroid of the coalesced droplet starts moving upward while the flexible substrate starts deforming under the force exerted by the coalescing droplets, as indicated by its local downward displacement. The deforming flexible substrate eventually causes the corresponding jumping droplet curve to deviate from that for the rigid substrate at ~3 ms.

We further investigate the effect of substrate flexibility on the droplet coalescence and jumping dynamics by performing droplet coalescence experiments on a series of thin, flexible superhydrophobic sheets of PDMS, copper, and aluminum of different lengths $L_0$, and for two coalescing droplet sizes of $D_0 \sim$ 1 mm and ~ 1.2 mm. As shown in Figure 2A, we find that the droplet jumping velocity $(v_j)$ reduces significantly with increase in the flexible substrate deformation $(y_j)$. Here, $y_j$ is measured at the moment of droplet jump and relative to the static deformation of the substrate due to the weight of the coalescing droplets as shown in Figure 2B (refer Supplementary Information Section S2 for further details).

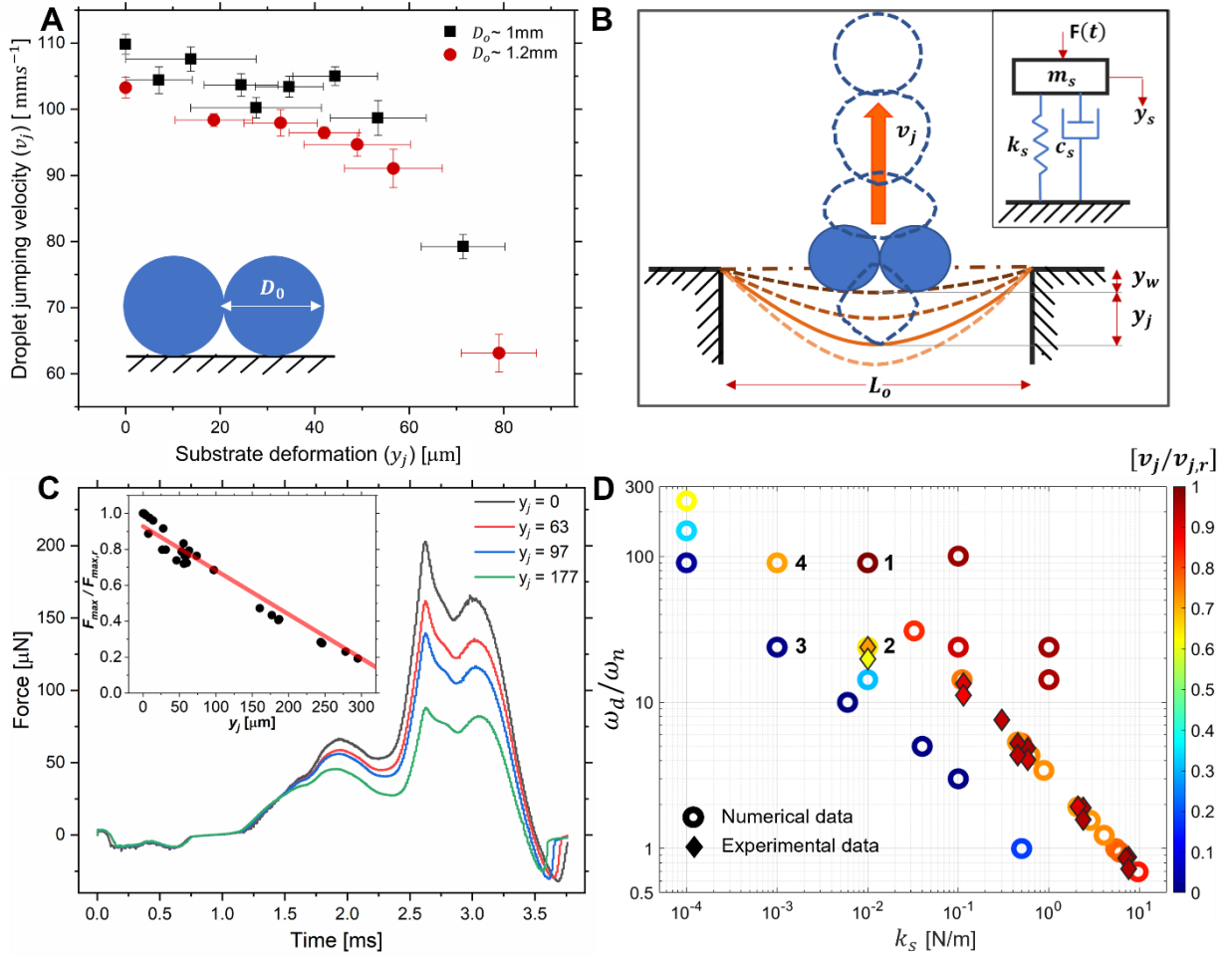

***Figure 2 A****: Coalescence induced droplet jumping velocity $(v_j)$ as a function of substrate deformation at the moment of droplet jump from the surface $(y_j)$. Results for two droplet sizes are shown. Zero value of deformation represents the case of the rigid substrate. Variation in substrate deformation is realized by varying the substrate length as well as the substrate material (refer Table S1 in Supplementary Information Section S3 for further details). **B**: Schematic (not to scale) defining substrate static deformation due to weight of the droplets $(y_w)$ and substrate deformation at the moment of jump $(y_j)$. **C**: Force exerted by the coalescing droplets $F(t)$ for various levels of $y_j$. Inset figure illustrates the relationship between the maximum coalescence force ratio $(F_{max}/F_{max,r})$ and $y_j$. Here, $F_{max}$ and $F_{max,r}$ represent the maximum coalescence force on flexible and rigid substrates, respectively. **D**: Ratio of the droplet jumping velocity on flexible and rigid substrate $(v_j/v_{j,r})$ as a function of substrate stiffness $(k_s)$ and frequency ratio $(\omega_d/\omega_n)$. Diamond markers indicate experimental results for the two droplet sizes corresponding to the results shown in 2A. Circular markers indicate numerical results.*

Figure 2C compares the computed force $F(t)$, from initiation of coalescence till the moment of jump, for rigid and three exemplar flexible substrates. It is evident that substrate flexibility significantly reduces the coalescence force, and this ultimately is responsible for the reduction of the jumping velocity of the droplet. This force $F(t)$ is estimated from a simplified fluid-structure interaction modeling framework wherein the droplet coalescence process is modeled through three-dimensional Volume-of-Fluid (VoF) method and the effect of substrate flexibility on the coalescence process is captured by using a single degree of freedom, linear spring-mass-damper system model (see inset of Figure 2B) (36). ((refer Materials and Methods and Supplementary Information Section S3 for further details of VoF-based fluid-structure interaction framework). The spring-mass-damper system model is governed by the following equation (35).

$$m_s\ddot{y}(t) + c_s\dot{y}(t) + k_s y(t) = F(t) \quad [1]$$

where $F(t)$ is the force exerted by the coalescing droplets on the substrate, $y(t)$ is the substrate deformation, and $m_s$, $k_s$ and $c_s$ are the substrate effective mass, stiffness, and damping coefficient, respectively. The damping coefficient for all the flexible substrates considered is found to be negligible, and the effective mass and stiffness of each substrate is estimated using modal analysis (refer Supporting Information Section S3 for details). We utilize this framework to obtain $F(t)$ on multiple flexible substrates with a wide range of $k_s$ and natural frequency $\left(\omega_n = \sqrt{\frac{k_s}{m_s}}\right)$ values. An analysis of the coalescence force $F(t)$ reveals that the maximum coalescence force $\left(F_{max} = \max_t[F(t)]\right)$ reduces linearly with $y_j$ (see inset Figure in 2C). Moreover, as shown in Figure 2C, the temporal variation of $F(t)$ is similar across most of the flexible substrates. Thus, $F(t)$ for any flexible substrate can be expressed as $F(t) \approx \left(1 - \frac{y_j}{y_0}\right) F_r(t)$ where $y_0$ is a fitting constant and $F_r(t)$ is the computed coalescence force on the rigid surface. These observations allow us to develop a semi-analytical model for the prediction of coalescence induced droplet jumping velocity and to isolate the fundamental parameters governing the droplet coalescence process on flexible substrates. Here, first $F_r(t)$ is expressed in terms of its dominating frequency components obtained through Fast Fourier Transform (FFT) analysis as $F_r(t) = \sum_i A_i \sin(\omega_i t) + B_i \cos(\omega_i t)$ (refer Supplementary Information Section S4 for details). Subsequently a solution of equation [1] yields the following relation for $y_j$

$$y_j = \frac{y_0 f_j}{y_0 k_s + f_j},$$

where $f_j = \sum_i \frac{1}{\left(1-\left(\frac{\omega_i}{\omega_n}\right)^2\right)} \left\{ A_i \left[\sin(\omega_i t_j) - \left(\frac{\omega_i}{\omega_n}\right)\sin(\omega_n t_j)\right] + B_i [\cos(\omega_i t_j) - \cos(\omega_n t_j)] \right\} \quad [2]$

and $t_j$ represents the moment of droplet jump. With $y_j$ known, the jumping velocity of the droplet for any substrate can be calculated based on the net force balance for the droplet as below.

$$v_j = \frac{1}{m_d} \int_0^{t_j} (F(t) - F_\sigma(t) - m_d g) dt$$

$$= \frac{1}{m_d}\left(1 - \frac{y_j}{y_0}\right) \sum_i \left\{\left(\frac{B_i}{\omega_i}\right) \sin(\omega_i t_j) - \left(\frac{A_i}{\omega_i}\right)[\cos(\omega_i t_j) - 1]\right\} - \left(v_0 - \frac{y_j}{t_0}\right) - g t_j \quad [3]$$

where $m_d$ is the total mass of the two coalescing droplets. The term $\frac{1}{m_d}\int_0^{t_j} F_\sigma(t)dt$ represents the contribution of the capillary adhesion between droplet and surface due to water-air interfacial surface tension. We find that it can be related to substrate deformation as $\frac{1}{m_d}\int_0^{t_j} F_\sigma(t)dt \approx \left(v_0 - \frac{y_j}{t_0}\right)$ where $v_0$ and $t_0$ are fitting constants (refer Supplementary Information Section S5 for further details).

The semi-analytical solution for jumping velocity given by equations [2-3] reveals that the jumping velocity for any substrate is a function of substrate stiffness $k_s$ and frequency ratios $\left(\frac{\omega_i}{\omega_n}\right)$ for the substrate. Here we also find that the dominating frequency, $\min_i[\omega(i)]$, is within ~15% of droplet oscillating frequency ($\omega_d = 2\pi/\tau_d$) based on the inertial-capillary time scale for the droplet $\left(\tau_d = \frac{\pi}{4}\sqrt{\frac{\rho D_0^3}{\sigma}}\right)$ (19, 35, 36). Here, $\rho$ and $\sigma$ are the liquid density and liquid-air interfacial surface tension respectively. Essentially, this semi-analytical model implies that the jumping velocity can be analyzed in terms of the frequency ratio ($\omega_d/\omega_n$) and substrate stiffness ($k_s$) (refer Supplementary Information Section S5 for further details). Figure 2D illustrates the experimental and numerical results for droplet coalescence induced jumping velocity ratio $(v_j/v_{j,r})$ as a function of these parameters. Here, $v_{j,r}$ is the jumping velocity on the rigid substrate. It is evident from the figure that the jumping velocity can be reduced by a combination of low ($k_s$) and low ($\omega_d/\omega_n$) values and a Pareto front of such combinations can be clearly seen in Figure 2D. This is because such substrates attain high deformation $y_j$ and thus minimize the reaction from substrate on the drop for coalescence induced jump (see Figure S6A).

This synergistic effect of substrate ($k_s$) and ($\omega_d/\omega_n$) can be understood by considering droplet coalescence on four flexible substrates marked in Figure 2D. Figure 3A illustrates the computed substrate deformation for these substrates. Although a substrate with a lower $k_s$ deforms more under the imposed coalescence force (e.g., compare substrates 1 and 4), reducing the $k_s$ alone is not sufficient to maximize the substrate deformation. This is because even if $k_s$ is low, the substrate with a relatively higher $\omega_d/\omega_n$ has higher inertia and consequently slower response to the force imposed by coalescing droplets (e.g. compare substrates 3 and 4). Essentially, this means that to maximize substrate deformation at the time of droplet departure and thus minimize droplet jumping velocity, a substrate with low $k_s$ as well as low $\omega_d/\omega_n$ is required, which, among these four substrates, is achieved for substrate 3. This is also reflected by the comparison of droplet upward kinetic energy as shown in Figure 3B. The maximum deformation for substrate 3 translates into the minimum conversion of excess surface energy to upward kinetic energy. Here, the normalized upward kinetic energy ($UKE^*$) at any time $t$ is calculated as $UKE^* = 0.25 m_d v_c^2(t)/(\Delta A \sigma)$, where $v_c(t)$ is the instantaneous droplet centroid velocity, and $\Delta A$ is the difference between total surface area of the two coalescing drops at the beginning coalescence (i.e. time $t = 0$) and at time $t_j$ (also see Figure S6B).

Figure 3C illustrates the evolution of drop shape and drop velocity field during droplet coalescence on a superhydrophobic rigid substrate, flexible substrate 3, and droplet coalescence in air. The neck formed during coalescence expands freely for drop coalescence in air. Consequently, coalescence of drops of the same size in air is symmetric, resulting in no movement of drop normal to the coalescence axis. In contrast, on a rigid substrate, the coalescence symmetry is broken, thus causing the drops to jump. On the flexible substrate 3 with low $k_s$ and low $\omega_d/\omega_n$ ratio, this symmetry breaking is minimized as the substrate deforms under the force exerted by the coalescing drops. The resulting velocity field and drop shape thus evolve closer to the case of drop coalescence in air.

The optimum combination of $k_s$ and $\omega_d/\omega_n$ ratio for a flexible substrate can be obtained through appropriate choice of the geometry and material of the substrate. In our experiments, we obtained a jumping velocity reduction of upto ~ 40% (i.e. $v_j/v_{j,r}$ ~ 0.6) for 1.2 mm droplets coalescing over a PDMS substrate of length, width, and thickness as 15mm, 4.2mm, and 52 μm (see Figure 2D). This translated into a nearly 64% reduction in droplet jumping height. Using our semi-analytical model, we can estimate that it is possible to design flexible substrates that can attain even higher levels of jumping velocity and height reduction through the appropriate choice of substrate material and geometry (refer Supplementary Information Section S7 for further details). For instance, the model predicts that a thin, flexible substrate of thickness, length, and width as $(0.03 \times 15 \times 4)$ mm and made from polymeric material, such as Low-density Polyethylene (LDPE), with density ~918 kg/m$^3$ (*37*, *38*) can attain a jumping velocity and height reduction as high as 70% ( i.e. $v_j/v_{j,r}$ ~ 0.3) and 90% respectively.

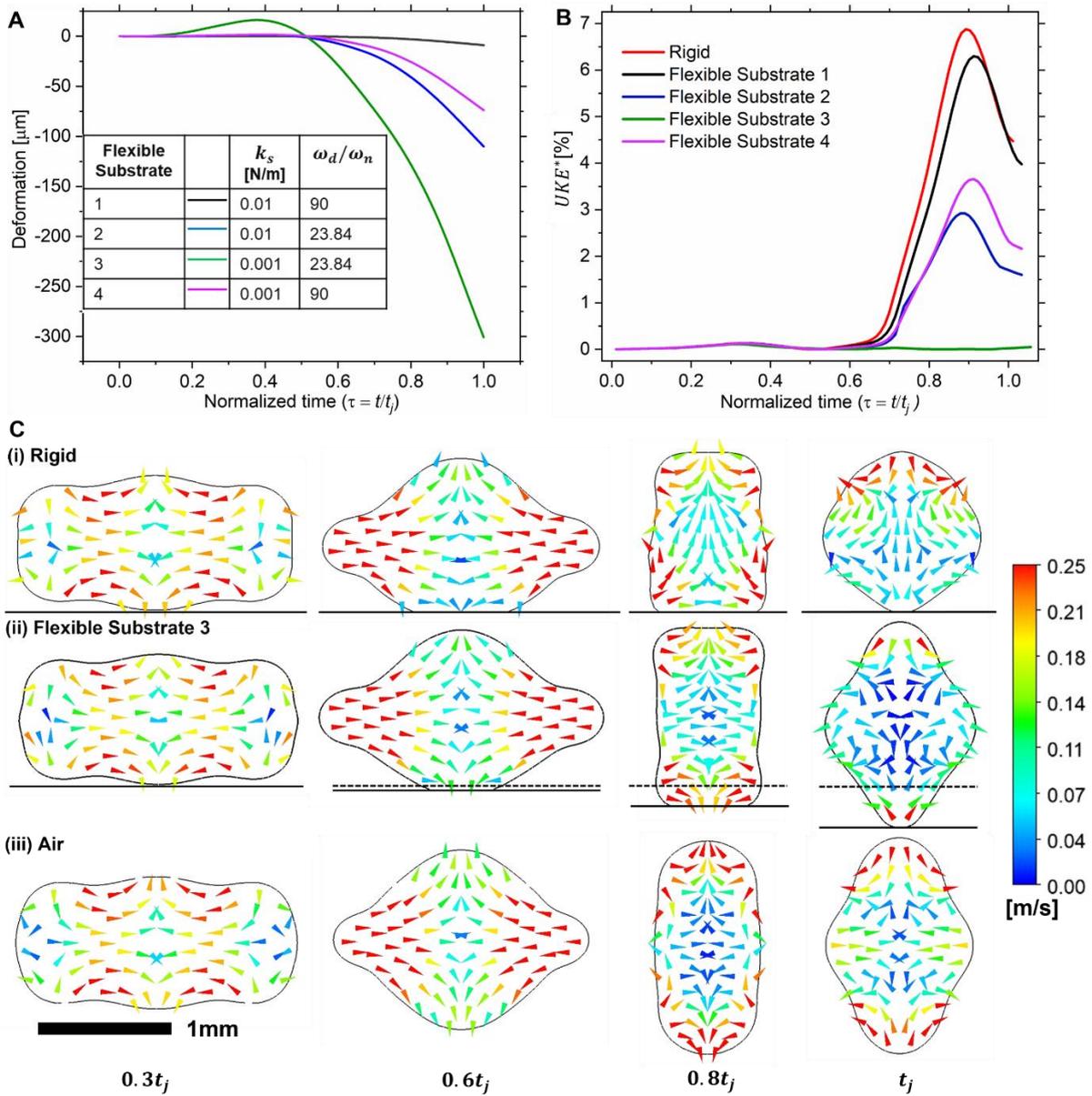

***Figure* 3A:** *Substrate deformation as a function of normalized time $\tau = t/t_j$, for the four substrates marked in Figure 2D. Substrate stiffness ($k_s$) and frequency ratios ($\omega_d/\omega_n$) are listed in the figure.* ***B****: Normalized upward kinetic energy of drop for the four substrates. Among the four substrates, substrate 3 with low $k_s$ and $\omega_d/\omega_n$ ratio attains the largest deformation and consequently results in lowest droplet upward kinetic energy.* ***C:*** *Comparison of computed droplet shape and velocity field for coalescence of drops with $D_0 = 1$ mm on i) rigid superhydrophobic substrate, ii) flexible superhydrophobic substrate 3 and iii) for droplet coalescence in air. Chosen time instants correspond to initiation of coalescence, $0.3t_j$ (neck contact with substrate), $0.6t_j$ (maximum elongated configuration of drop), $0.8t_j$ (maximum contact area between drop and substrate) and $t_j$ (moment of droplet jump). For droplet coalescence in air, snapshots at the same time instants are shown. The solid and the dotted black line in (i) and (ii) represents the initial and instantaneous position of the surface.*

Subsequently, we explore the effect of substrate flexibility on coalescence induced jumping of droplets that are significantly smaller than the ones considered in the controlled experiments discussed above. This is especially relevant to water harvesting from dew and fog wherein a superhydrophobic surface can induce jumping departure of droplets, thus achieving high rates of surface renewal. However, such jumping droplets are vulnerable to loss through entrainment in surrounding air flow. The reduced jumping velocity on a flexible superhydrophobic surface can be leveraged to reduce such loss. Here we investigate this aspect through a controlled fog harvesting experiment wherein the fog is intercepted by a thin superhydrophobic flexible substrate. Typical fog droplets lie in the size range of ~ 3-50 µm, as reported by multiple studies (*39*, *40*). Compared to millimetric droplets, such small droplets exert a much smaller level of force on the substrate during coalescence. However, the coalescence dynamics for smaller droplets are similar to larger droplets, as verified by the evolution of force shown in Figure S8. Thus, in order to influence the jumping dynamics of such small droplets, a thin flexible superhydrophobic copper foil mounted in a vertical cantilever configuration is used. This orientation enables us to realize substrate stiffness $k_s$ that is nearly two orders of magnitude smaller than the stiffness of the horizontal fixed-fixed beam configuration. Additionally, to reduce the substrate $\omega_d/\omega_n$ ratio, the sample mass is reduced by removing excess material. A fog generator where generated droplets are of the size of around 6 microns (*41*) is used, and the droplet jumping velocity is characterized by recording the position of droplets falling from the substrate on a mirror finished aluminum collector sheet (refer Supplementary Information Section S9 for experimental details). Figure 4A and B show the stacked map of all drops falling from rigid (red) and flexible substrate (green) during fog exposure for over 2 hours. Each map consists of droplets of size ~ 100 microns, which is nearly two orders of magnitude larger than the fog droplets impinging on the substrate. Thus, the fog droplets intercepted by the substrate are departing from the surface predominantly because of the coalescence-induced jumping.

Figure 4C illustrates the normalized cumulative droplet distribution corresponding to the maps shown in Figure 4A. It is observed that for flexible surface, a large fraction of droplets (~80%) fall within a distance of 20 mm from the substrate, whereas, for the rigid substrate, only ~40% of the droplets fall within the same distance (refer Figure S10 for corresponding droplet distributions). These results illustrate the potential of using flexible substrates towards reducing the scattering of jumped droplets during fog and dew harvesting.

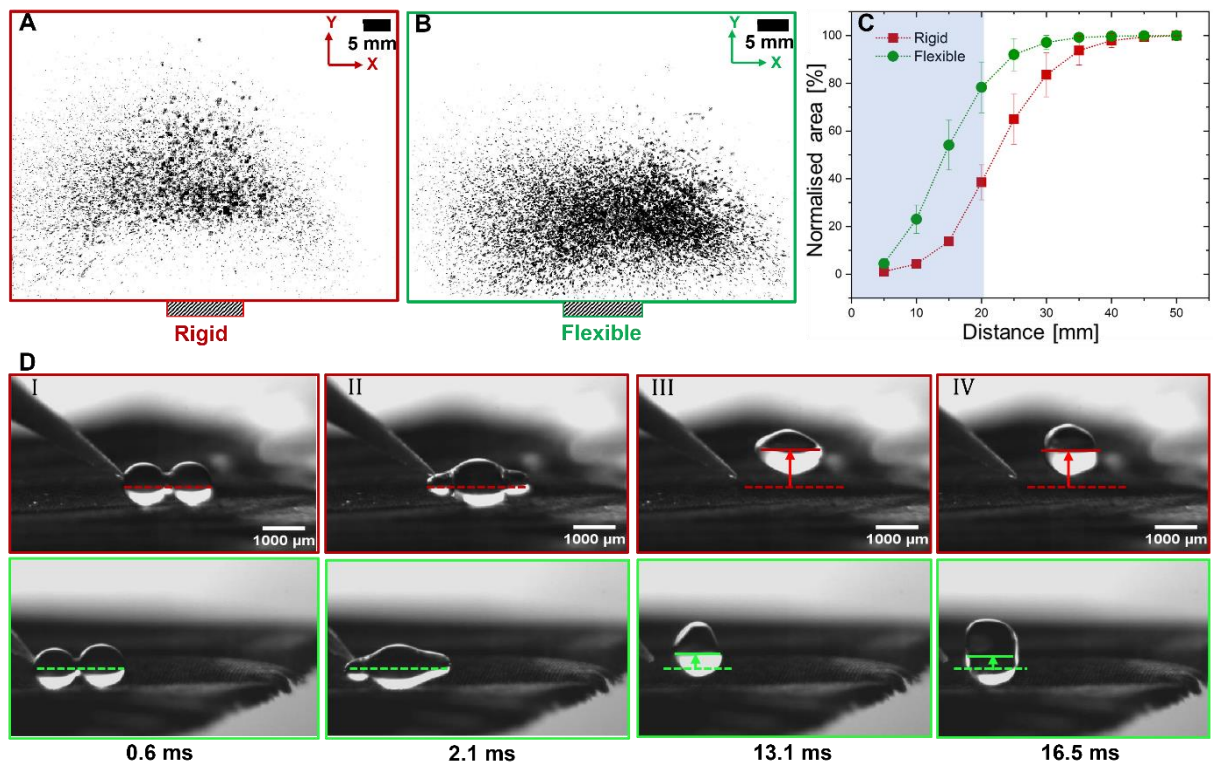

*Figure 4* Map of fog water droplets jumping from rigid (**A**) and flexible (**B**) superhydrophobic surfaces and deposited on a collector plate positioned as shown in Figure S9. The substrate is 13 mm wide and positioned as shown by the hatched box in both cases. **C**: Comparison of normalized cumulative droplet distribution in terms of normalized cumulative coverage area on the collector plate and as a function of distance along Y from substrate for rigid (red) and flexible (green) substrates. The shaded region marks region on the collector plate, which contributes ~80% of the total droplet coverage area for the flexible substrate. **D**: Image sequence of coalescence-induced droplet jumping on supported (red) vs. unsupported (green) butterfly wings (Papilio Polytes). Frame (a) represents the time just after coalescence events start. Subsequent frames are for maximum elongation of drop after the coalescence, maximum jump for flexible and rigid cases, respectively. The horizontal dashed lines in both image sequences indicate initial droplet position, and the vertical solid lines are guide for differentiating jumping height visually.

Superhydrophobic surfaces are also commonly observed in nature wherein many species have evolved self-cleaning ability that is critical to their survival. In many cases, this ability realized by coalescence induced droplet jumping helps insects such as cicadas (*42*) and water striders (*43*) to keep their wings and legs clean and free from moisture to maintain their desired functions. Similarly, there are a few examples of plants with superhydrophobic self-cleaning surfaces, such as lotus (*44*) and wheat leaves (*45*). Since many of such natural superhydrophobic surfaces are also flexible, substrate flexibility is likely to influence the overall jumping dynamics. To investigate this aspect, we perform droplet coalescence experiments on a wing of Papilio Polytes, a common butterfly of our region (*46*). Two sets of experiments are performed, one mimicking the natural condition, where the butterfly wing is free to move like a cantilever, and the other wherein the wing is backed with rigid support to simulate the condition of a rigid superhydrophobic substrate. Figure 5D compares coalescence induced droplet jumping on the butterfly wing under supported (top row marked by red) and natural (bottom row marked by green) states. In this experiment, the wing flexibility reduces the jumping height by ~ 45% compared to when the wing is supported by a rigid substrate. Thus the coalescence induced droplet jumping is significantly subdued on the wing under its natural condition due to its inherent flexibility. Thus, we speculate that the removal of dew and fog droplets through coalescence induced jumping, and consequently the self-cleaning ability, can be significantly affected by the inherent stiffness of the natural surface.

In summary, our study provides fundamental insights into the phenomenon of coalescence induced droplet jumping on flexible superhydrophobic substrates. Our experiments and numerical investigations reveal that coalescence induced droplet jumping velocity can be significantly curtailed through synergistic combinations of low substrate stiffness and substrate frequency. Further, the numerical modeling framework and the semi-analytical model can be used to design flexible superhydrophobic substrates for use in applications such as dew and fog harvesting. We present a proof-of-concept demonstration on gainfully exploiting this aspect in the context of fog harvesting. Lastly, through droplet-droplet coalescence experiments on an exemplar natural surface, we demonstrate that this reduced droplet jumping effect is likely to manifest on natural superhydrophobic substrates, many of which are flexible.

**<u>Materials and Methods</u>**

**Super-hydrophobic surface fabrication:** We fabricated rigid and flexible super-hydrophobic substrates by spray coating of commercially available solution of silica nanoparticles, homogenously dispersed in isopropanol solution (Glacco Mirror coat "Zero". Soft 99 Co.) (*19*). The rigid super-hydrophobic surface was fabricated on a glass substrate. For the flexible substrates, we used copper foil (Nanoshel LLC) of 10-micron thickness, aluminum foil of 20-micron thickness (Nanoshel LLC), and PDMS (Sylgard 184 Dow Corning). For preparing the PDMS substrate, a mixer of base and crosslinker in stoichiometric ratio of 10:1 was spin-coated on the glass substrate. Subsequently, PDMS was cured for 2 hours at $80^o$ C in an oven (Binder Inc.), and the cured PDMS was peeled off from the glass substrate to obtain a thin PDMS sheet. PDMS substrates of various thicknesses were prepared by varying the spin coating speed.

**Wettability Characterizations:** We measured apparent advancing ($\theta_a$) and receding contact angles ($\theta_r$) for all the surfaces using a contact angle meter setup (Holmarc Opto-Mechatronics Ltd, Model No- HO-IAD-CAM-01A). For superhydrophobic copper and aluminum substrate, the apparent advancing contact angle and contact angle hysteresis is $163 \pm 1.2^0$ and $3.8 \pm 1.3^0$ respectively. For PDMS substrate, apparent advancing contact angle and contact angle hysteresis is $161 \pm 2.1^0$ and $4.2 \pm 1.2^0$ respectively. For superhydrophobic glass substrate, the apparent advancing contact angle and contact angle hysteresis is $163 \pm 2.4^0$ and $3.4 \pm 1.4^0$ respectively.

**Experimental procedure:** We performed droplet-droplet coalescence experiments on rigid and flexible superhydrophobic substrates. The stiffness ($k_s$), and the natural frequency ($\omega_n$) of the substrate was varied by choosing various combinations of substrate material (aluminum, copper and PDMS) and dimensions. The substrate was carefully mounted in a fixed-fixed beam configuration where one end of substrate was fixed to a holder with double-sided tape and the other end was fixed by placing a dead load of 10 μN. This was done to avoid any pre-stretching of the thin substrates, as any resulting pre-stress can cause a significant change in stiffness and natural frequency of the thin flexible substrate. All the experiments were carried out on an actively vibration-isolated optical table to nullify the effect of outside vibrations on the experiments. The substrate was grounded to avoid any effect of static charge on the process (*47*) (refer to Figure S1 for details on the experimental setup). We used super-hydrophobic micropipette tips developed in-house to dispense the droplets on the substrate. The superhydrophopic micropipette tips were fabricated by multiple cycles of dip coating in a particle-polymer solution followed by heating in the oven at $80^oC$. The polymer solution is prepared by mixing 17 ml of acetone from Merck, 3 ml of Capstone ST 200 and one gram of fumed hydrophobic silica nano particle from Evonik (*34*). In each experiment, two drops were first dispensed on the substrate and subsequently one drop was slowly moved and brought in contact with the other drop using a super-hydrophobic wooden tip to trigger coalescence. We ensured through image analysis that any kinetic energy imparted to the droplet in this process was minimal (< ~5% of surface energy of the droplet) (*48*). All the images were recorded at 10000 frames per second using Photron Fastcam SA4 camera. After the experiments, the images were analyzed for droplet movement and substrate deformation using Fiji ImageJ. (Refer Section S2 for further details**).**

**Three-dimensional numerical modeling of coalescence-induced droplet jumping:**

The droplet coalescence process is modeled based on the three-dimensional Volume-of-Fluid method. The governing equations are solved with the commercial software ANSYS Fluent (*11, 13*). We have chosen an incompressible, laminar flow model for our simulation and the contact angle on the surface is assumed to be $180^0$. This assumption is valid due to the high advancing contact angle and very low contact angle hysteresis on the fabricated surfaces (*11, 49, 50*). The symmetry of the coalescence process is utilized to reduce the overall computational domain wherein one half of a coalescing droplet is modeled (*13*). The surface on which droplets are lying is modeled as a no-slip wall boundary. Open boundary condition is imposed on all the other sides of the domain. The domain size chosen in our simulation is 4R×4R×4R where R is droplet radius (*13*) (see Figure S3). The numerical model for the rigid substrate is validated with the result from Liu et al. (*50*). For the case of flexible substrates, the surface is modeled as a no-slip moving wall. The wall is moved according to the substrate deformation computed in each timestep through a user-defined function. The function calculates the substrate deformation by modeling the substrate as a spring-mass system according to Equation [S1] (*51, 52*). The force $F(t)$ at any time $t$ is obtained by integrating the fluid pressure on the wall boundary. The numerical model for droplet coalescence on a flexible substrate is validated against experimental results (refer supplementary information section S3 for further details).

**Natural Surface:** Papilio Polytes or the Common Mormon butterflies were collected during spring season from the region adjoining Indian Institute of Technology Ropar. Droplet coalescence experiments on the wing were performed under two conditions, by mounting the wing in a horizontal cantilever configuration to mimic the natural condition and by supporting the wing by a glass slide using a thin layer of water (*35*). The advancing and the receding contact angle on the wings were measured as $159\pm4.2^O$ and $153\pm3.3^{O,}$ respectively.

**Author Affiliations**

Gopal Chandra Pal[†], Siddharth SS[†], Chander Shekhar Sharma[†]

[†]Thermofluidics Research Lab, [‡]Department of Mechanical Engineering, Indian Institute of Technology Ropar, Rupnagar, Punjab 140 001, India
*Email: chander.sharma@iitrpr.ac.in, Ph: +91-1881-232358

Manish Agarwal[‡]

[‡]Department of Mechanical Engineering, Indian Institute of Technology Ropar, Rupnagar, Punjab 140 001, India
**Author Affiliations**

Gopal Chandra Pal[†], Siddharth SS[†], Chander Shekhar Sharma[†]

[†]Thermofluidics Research Lab, [‡]Department of Mechanical Engineering, Indian Institute of Technology Ropar, Rupnagar, Punjab 140 001, India
*Email: chander.sharma@iitrpr.ac.in, Ph: +91-1881-232358

Manish Agarwal[‡]

[‡]Department of Mechanical Engineering, Indian Institute of Technology Ropar, Rupnagar, Punjab 140 001, India

## Author Contributions

C.S.S. conceived and supervised the research. C.S.S. and G.C.P. designed the experiments. G.C.P. fabricated the samples, performed the experiments and conducted the numerical simulations. M.A., G.C.P and S.S.S. developed the simplified fluid-structure interaction modeling framework. G.C.P, C.S.S. and M.A. analyzed the results. M.A. and C.S.S. developed the analytical model. G.C.P, C.S.S. and M.A. wrote the paper which was read and approved by all authors.

## Acknowledgments

We thank Ravi Kant Upadhyay and Anand S. for helpful discussions on sample fabrication and experimental methods. C.S.S. acknowledges the funding for this work from Indian Institute of Technology Ropar through ISIRD scheme (Grant No. 9-388/2018/IITRPR/3335).

# Controlled coalescence-induced droplet jumping on flexible superhydrophobic substrates

Gopal Chandra Pal[†], Siddharth SS[†], Manish Agarwal[‡], Chander Shekhar Sharma[†]*

[†]Thermofluidics Research Lab, [‡]Department of Mechanical Engineering, Indian Institute of Technology Ropar, Rupnagar, Punjab 140 001, India
*Email: chander.sharma@iitrpr.ac.in, Ph: +91-1881-232358

## Table of contents

| S1 | Experimental setup for droplet-droplet coalescence experiments on rigid and flexible substrates |
|---|---|
| S2 | Image processing for droplet jumping height measurement and error analysis |
| S3 | Three-dimensional numerical modelling of coalescence-induced droplet jumping: Model validation |
| S4 | Fast Fourier Transform (FFT) analysis of force exerted by the coalescing droplets on the substrate |
| S5 | Semi-empirical model for estimation of jumping velocity |
| S6 | Details on substrate deformation and energy analysis |
| S7 | Optimization of substrate geometry and material properties for minimization of jumping velocity. |
| S8 | Effect of droplet size on coalescence dynamics |
| S9 | Experimental details for fog harvesting |

# S1. Experimental setup for droplet-droplet coalescence experiments on rigid and flexible substrates

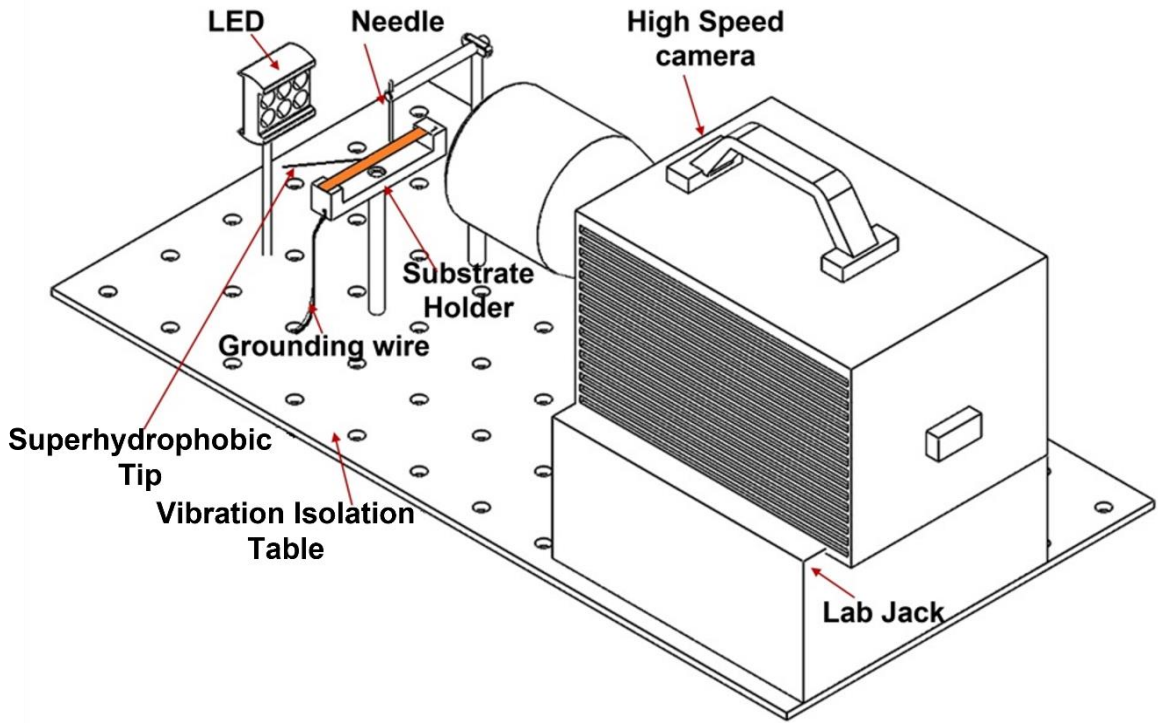

*Figure S1*: Experimental setup for coalescence induced droplet jumping

# S2: Image processing for droplet jumping height measurement and error analysis

a) <u>Image processing</u>: The images taken by the high-speed camera are processed for measuring the jumping height in both the cases of the rigid and flexible substrate. The flexible substrate deforms by a finite amount ($y_w$) from the horizontal position (marked by the dotted line) due to the self-weight of the droplets dispensed on it, as shown in Figure S2a. The substrate deformation due to the force exerted by the coalescence process is measured with respect to this static deformed configuration of the substrate as highlighted by Figure 2B in the main text.

All the experimental images are converted to a binary image through thresholding by using the ImageJ software. After conversion, the droplet motion is analyzed by measuring the movement of the droplet centroid over time. The resulting data is used to estimate the jumping velocity as $v_j \approx \sqrt{2gH_j}$ where $H_j$ is the vertical distance between droplet position at the moment of jump and at the maximum height from the surface as shown in Figure S2b. We also estimate $v_j$ by fitting a parabolic curve to the droplet trajectory obtained above (*1*). Both the estimates yield similar values of $v_j$ with a deviation of less than ~3% . Similarly, we also process the images to measure the substrate deformation due to the force exerted by the coalescence event.

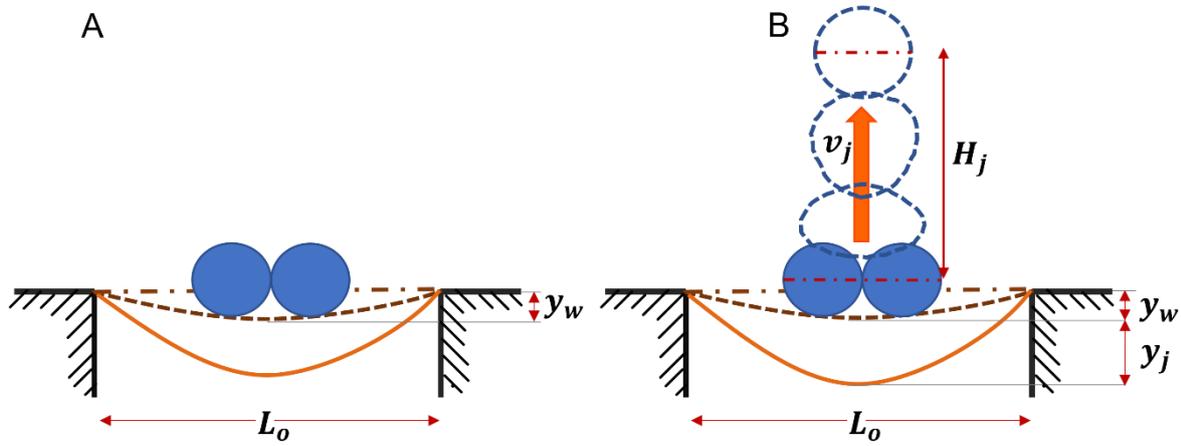

*Figure S2: (a) Substrate deformation due to self-weight and droplet weight (b) Measurement of $H_j$ for estimation of the jumping velocity $v_j$*

b) Error analysis:

i) Effect of experimental image processing on droplet movement measurement: When the droplet path as obtained from experimental images is analyzed, a path as shown in Figure 1B in main text is obtained. This undulating path is obtained due to the fact that the droplet centroid location measurement in each image is based on two-dimensional images of the three-dimensional jumping droplet. This aspect contributes towards experimental uncertainty in droplet trajectory which we estimate through multiple repeated trials of each experiment. Another source of uncertainty in centroid location arises from the thresholding of images due to the finite resolution of optical imaging. Hence the net error in centroid location at any time instant is obtained through error propagation as $\Delta E = \sqrt{\Delta E_s^2 + \Delta E_R^2}$, where $\Delta E_R$ is the random error estimated from multiple repeated trials of an experiment and $\Delta E_s$ is the systematic error due to finite resolution of image processing (*2*).

ii) Effect of droplet dispensing through superhydrophobic micropipette tip: During the coalescence experiments, droplets are dispensed on the substrate using a superhydrophobic tip. We find that there is a variation of less than 4% in terms of droplet diameter in all the repeated trials for particular cases of coalescence in flexible/rigid substrate. We have performed a minimum of three trials for all the cases.

## S3: Three-dimensional numerical modelling of coalescence-induced droplet jumping: Model validation

a) Calculation of natural frequency, stiffness, equivalent mass and damping coefficient of the flexible substrate:

The substrate is modeled as a beam mounted in a fixed-fixed configuration as in experiments. The natural frequency ($\omega_n$) and modal stiffness ($k_s$) of the substrate is obtained by performing modal analysis in the ANSYS Mechanical software (*3*). Here, the governing equation for the modal analysis is given by equation (S1).

$$(\mathbf{K} - \omega_n^2 \mathbf{M})\phi = 0 \quad (S1)$$

where $\mathbf{K}$, $\mathbf{M}$ and $\phi$ represent stiffness matrix, mass matrix and mode shape respectively. Grid independence is performed to ensure the correctness of the obtained model solution. The material properties of copper and aluminum are taken from the substrate manufacturer. For PDMS, the material properties are obtained from literature (*4*). The stiffness ($k_s$) and natural frequency ($\omega_n$) corresponding to mode 1 for various substrates are listed in Table S1.

The damping ratio for aluminum and copper is taken from the literature and the values found to be approximately ~ $10^{-3}$(*5, 6*) and for PDMS substrate the damping ratio is estimated as ~ $10^{-2}$ using logarithmic decrement method where $\zeta = (ln|\delta_1/\delta_2|)/2\pi$. $ln|\delta_1/\delta_2|$ represents the logarithmic decrement (*7*). Thus, we have neglected the effect of damping in our analysis.

| S. No. | Material | Length, mm | Width, mm | Thickness, micron | $\omega_n$, rad/sec | $k_s$, N/m |
|---|---|---|---|---|---|---|
| 1 | Aluminum | 25 | 4 | 20 | 1045.64 | 2.4 |
| 2 | Aluminum | 40 | 4 | 20 | 408.2 | 0.587 |
| 3 | Copper | 25 | 4 | 10 | 376.21 | 0.45 |
| 4 | Copper | 40 | 4 | 10 | 146.95 | 0.1137 |
| 5 | PDMS | 15 | 4.4 | 52 | 83.05 | 0.01 |

*Table S1*: Substrate natural frequency and stiffness values corresponding to mode 1

b) Droplet coalescence on rigid superhydrophobic surface: To capture the coalescence-induced droplet jumping event, we perform a 3D VOF numerical simulation of two droplets resting on a smooth super-hydrophobic surface using commercial software ANSYS Fluent. For grid independence, we consider three grids with 26 elements per radius, 40 elements per radius, and 53 elements per radius respectively for a droplet size of 1 mm droplet diameter. The deviation in force magnitude in the case of 40 and 53 elements per radius is lower than that for the case of 26 elements and 40 elements per radius. Moreover, the maximum force magnitude differs by less than 2.5 % and the droplet detachment time differs by less than 5% for the meshes with 40 and 53 elements per radius. Thus, we have used 40 elements per radius for all the numerical simulations. The computational domain with boundary conditions and the computational mesh is shown in Figure S3A. With the mesh density fixed, the numerical result for coalescence of droplets with diameter of 1.06 mm is validated with the corresponding experimental result on a rigid superhydrophobic surface as shown in Figure S3B.

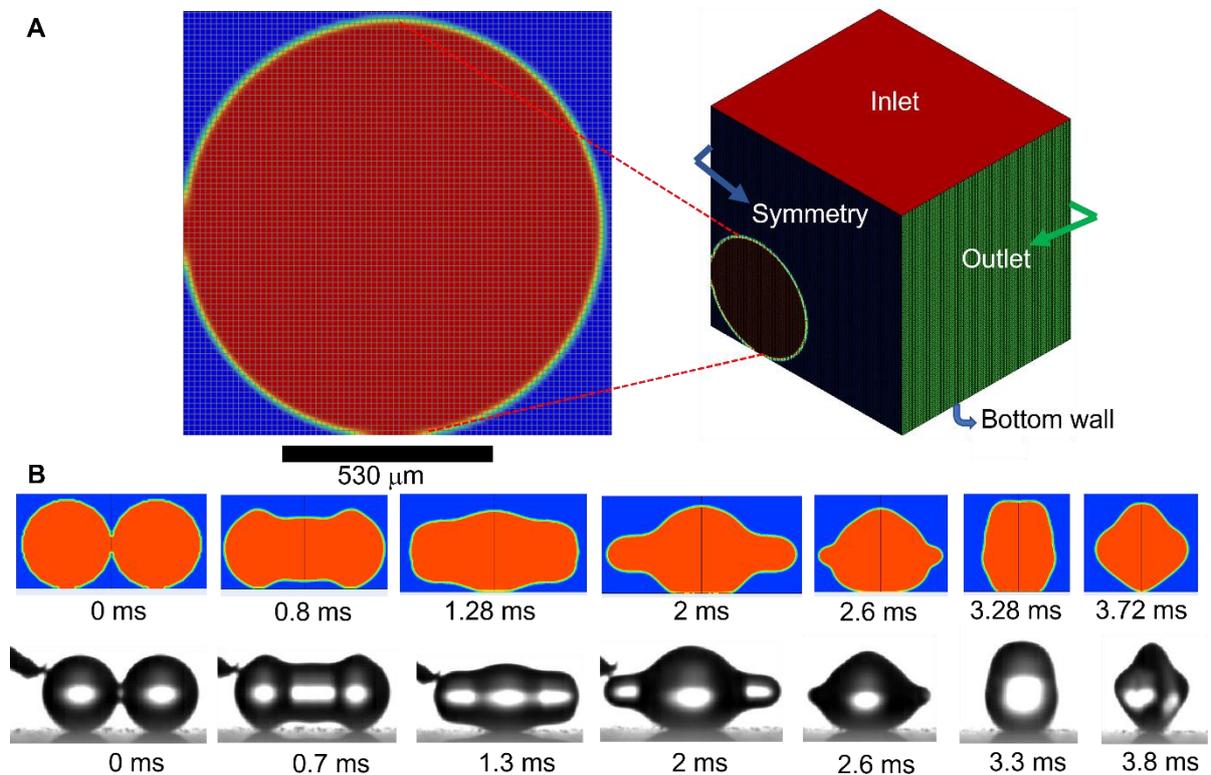

*Figure S3 A: Computational mesh and boundary conditions for three-dimensional VoF based CFD simulation B: Validation of numerical result bottom row with the experimental result (top row) for a droplet size $D_0 = 1.06$ mm*

c) Droplet coalescence on flexible superhydrophobic surface:

For modeling droplet coalescence on flexible superhydrophobic substrate, in addition to the above described VoF modeling of coalescing droplets, we need to also model the substrate deformation under the effect of force exerted by coalescing droplets and its effect on the coalescence process. To this effect, we adopt a simplified fluid-structure interaction modeling approach where the substrate deformation is realized through a dynamic moving boundary strategy (*7, 8*) Here, the motion of the wall representing the substrate is defined based on the substrate deformation calculated by the lumped spring mass system that can account for multiple vibration modes of the substrate(*9*). We have taken the assumption of linearity while modeling the substrate as the lumped system. Although the substrate deformation is large and thus the stiffness is inherently non-linear, we find that the linear lumped spring-mass system model still sufficiently captures the substrate deformation. The lumped mass system is solved using the forward Euler method for every iteration using a user-defined function. The calculated substrate velocity is passed to the solver for the wall motion while the governing equations of mass conservation, momentum and fluid fraction are solved by the pressure solver (*10*).

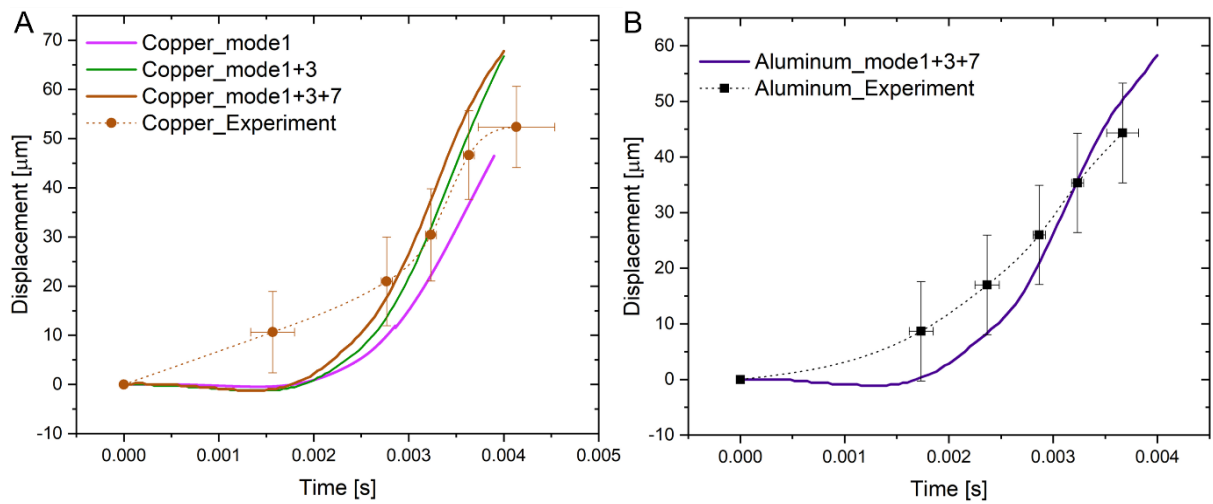

*Figure S4: Comparison of substrate displacement obtained from three-dimensional VoF method based simplified fluid-structure interaction modeling framework, by including various number of substrate deformation modes, with the experimental measurements for coalescence of 1.06 mm drops on A: thin Copper substrate and B: Aluminum substrate of 40 mm length.*

Modal analysis as discussed in subsection (a) above is used to obtain modal stiffness and modal mass values for multiple modes. This enables us to isolate the dominating modes for substrate deformation under the force exerted by coalescing droplets, by comparing the predicted substrate deformation against experimental measurements. The experimental substrate deformation values are obtained through high-speed imaging. For instance, Figure S4A shows the comparison of substrate deformation obtained from experimental measurement and CFD simulations for a thin copper substrate of 40 mm length (Substrate 4 in Table S1). Three curves are shown for computational results that are obtained by including increasing numbers of vibration modes in the overall calculation. A comparison of predicted and measured substrate deformation shows that accounting for first, third and seventh vibration mode is sufficient for prediction of substrate deformation. This is also validated by the comparison of predicted and measured substrate deformation for a 40 mm long Aluminum substrate (Substrate 2 in Table S1), as shown in Figure S4b. Additionally, we noticed that the first mode accounts for nearly 75% of the substrate deformation in copper and 90% for the aluminum case during the droplet coalescence and jumping departure from the substrate. Thus, in order to develop a qualitative understanding of the effect of substrate flexibility parameters on droplet coalescence process, we considered only the first mode while calculating the substrate deformation in the user defined function and the same modeling approach was adopted for all the computational results shown in Figure 2D in main text.

**S4: Fast Fourier Transform (FFT) analysis of force exerted by the coalescing droplets on the substrate**

An FFT analysis has been performed in MATLAB to ascertain the leading harmonic components in the force exerted by coalescing droplets on rigid substrate $F_r(t)$. The resulting FFT components for the force applied on the rigid substrate are given in the below table. Also as stated in the main text, the transient is force has been computed from the CFD analysis. We find that the dominating frequency $[\omega_1]$ is within ~15% of droplet oscillating frequency $\left(\omega_d = \frac{2\pi}{\tau_d}\right)$ based on the inertial-capillary time scale for the droplet $\left(\tau_d = \frac{\pi}{4}\sqrt{\frac{\rho D_0^3}{\sigma}}\right)$ (9, 11, 12). Here, $\rho$ and $\sigma$ are the liquid density and liquid-air interfacial surface tension respectively.

| Frequency (Hz) | FFT Components (μm) |
|---|---|
| 0 | 0.09145 |
| 268.8172 | $-0.01363 + i\,0.1318$ |
| 537.6344 | $-0.0625 + i\,0.0187$ |

*Table S2*: *The FFT components for the transient force applied on the rigid substrate by droplet coalescence of $D_0 = 1\ mm$*

**S5. Semi-empirical model for estimation of jumping velocity**

The deformation of the flexible substrate due to the force exerted by coalescing droplets is governed by equation [1] in main text and reproduced here for reference

$$m_s \ddot{y}(t) + c_s \dot{y}(t) + k_s y(t) = F(t) \qquad [1]$$

where, $F(t)$ is the force exerted by the coalescing droplets on the substrate, and $m_s$, $k_s$ and $c_s$ are the substrate effective mass, stiffness and damping coefficient respectively. We have considered only one mode for structural deformation, since as shown in section S3, this is a reasonable assumption. Further, since the substrate has negligible damping, equation [1] reduces to

$$m_s \ddot{y}(t) + k_s y(t) = F(t). \qquad [S2]$$

Also as shown in the inset figure in 2C the force on the substrate is a linear function of the substrate deformation at time of departure $y_j$. Thus, the force exerted on the flexible substrate can be related to the rigid substrate case by

$$F(t) \approx \left(1 - \frac{y_j}{y_0}\right) F_r(t). \qquad [S3]$$

Further, the force on the rigid substrate is expressed in terms of its dominating frequency components (obtained in section S4) as:

$$F_r(t) = \sum_i A_i \sin(\omega_i t) + B_i \cos(\omega_i t) \qquad [S4]$$

Using equations [S3] and [S4], equation [S2] is solved for $y(t)$ with the following initial condition:

$$y = 0 \text{ and } \dot{y} = 0 \text{ at } t = 0 \qquad [S5]$$

Thus, a solution for $y(t)$ is obtained as below:

$$y(t) = \frac{1}{k_s}\left(1 - \frac{y_j}{y_0}\right) \sum_i \frac{1}{\left(1-\left(\frac{\omega_i}{\omega_n}\right)^2\right)} \left\{ A_i \left[\sin(\omega_i t) - \left(\frac{\omega_i}{\omega_n}\right)\sin(\omega_n t)\right] + B_i[\cos(\omega_i t) - \cos(\omega_n t)] \right\}$$

$$[S6]$$

Since $y = y_j$ at $t = t_j$, [S6] provides the solution for $y_j$ as shown in equation [2] in main text and reproduced here:

$$y_j = \frac{y_0 f_j}{y_0 k_s + f_j},$$

where $f_j = \sum_i \frac{1}{\left(1-\left(\frac{\omega_i}{\omega_n}\right)^2\right)} \left\{ A_i \left[\sin(\omega_i t_j) - \left(\frac{\omega_i}{\omega_n}\right)\sin(\omega_n t_j)\right] + B_i[\cos(\omega_i t_j) - \cos(\omega_n t_j)] \right\} \qquad [2]$

Thus, using [S3] and [S4], the force on flexible substrate is obtained as:

$$F(t) \approx \left(\frac{y_0 k_s}{y_0 k_s + f_j}\right) \sum_i A_i \sin(\omega_i t) + B_i \cos(\omega_i t) \qquad [S7]$$

The droplet jumping velocity can be obtained through force balance on the drop as:

$$v_j = \frac{1}{m_d}\int_0^{t_j}(F(t) - F_\sigma(t) - m_d g)dt$$

Using [S7] and the approximation $\frac{1}{m_d}\int_0^{t_j} F_\sigma(t)dt \approx \left(v_0 - \frac{y_j}{t_0}\right)$,

$$v_j = \frac{1}{m_d}\left(1 - \frac{y_j}{y_0}\right)\sum_i\left\{\left(\frac{B_i}{\omega_i}\right)\sin(\omega_i t_j) - \left(\frac{A_i}{\omega_i}\right)[\cos(\omega_i t_j) - 1]\right\} - \left(v_0 - \frac{y_j}{t_0}\right) - gt_j \quad [3]$$

with $y_j$ given by equation [2].

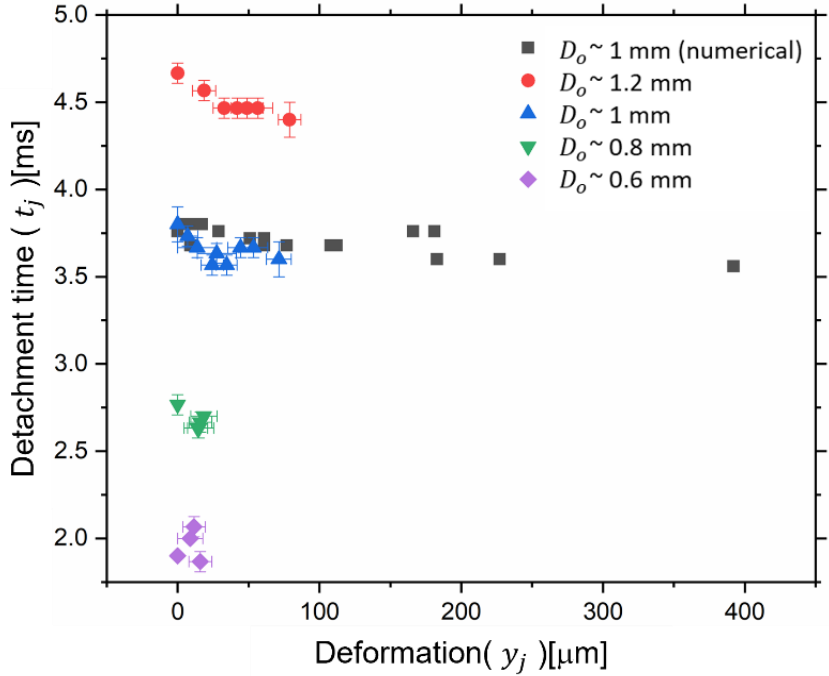

***Figure S5:*** *Droplet jumping time $t_j$ as a function of droplet size $D_0$ and substrate deformation $y_j$*

Here, $t_j$ is assumed to be invariant with substrate stiffness $k_s$ and frequency ratios $\frac{\omega_i}{\omega_n}$. This is because, as shown in Figure S5, $t_j$ is found to be dependent only on the size of the coalescing droplets $D_0$ and nearly independent of substrate deformation $y_j$. The above model is validated against results of the three-dimensional CFD model as shown in Figure S7A. The semi-empirical model successfully captures the overall trend for jumping velocity ratio as predicted by the more elaborate three-dimensional fluid-structure interaction modelling framework. Thus, the model can be used for comparative analysis of coalescence induced jumping velocity across flexible substrates of varying substrate stiffness and frequency ratios as shown in Figure S7C.

## S6. Details on substrate deformation and energy analysis

Figure S6A shows substrate deformation $y_j$ as a function of substrate stiffness ($k_s$) and frequency ratio ($\omega_d/\omega_n$). Experimental as well as numerical data is displayed. It is evident that $y_j$ is maximized for optimal combinations of low ($k_s$) and low ($\omega_d/\omega_n$) values. Figure S6B illustrates how substrate deformation $y_j$ affects the droplet upward kinetic energy ($UKE^*$) as a fraction of the total initial excess surface energy available. As $y_j$ increases, droplet upward kinetic energy decreases.

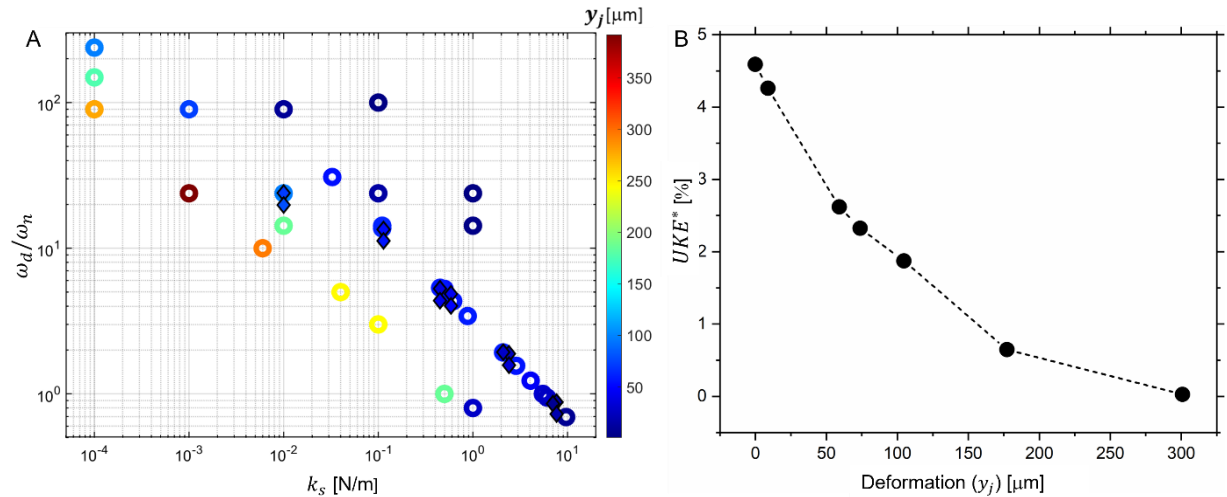

***Figure S6 A:*** *Substrate deformation $y_j$ as a function of $k_s$ and $\omega_d/\omega_n$. Diamond and circular markers correspond to experimental and numerical data respectively.* ***B:*** *Normalised droplet outward kinetic energy from initiation of coalescence till the moment of droplet jump, as a function of substrate deformation $y_j$*

## S7. Optimization of substrate geometry and material properties for minimization of jumping velocity.

For a fixed-fixed beam configuration, the substrate geometry for a given material can be optimized for minimization of jumping velocity by, for instance, varying the length of the substrate for fixed width and thickness. Figure S7A shows the effect of substrate length on jumping velocity ratio as predicted by three-dimensional CFD simulations (red dots). It is evident that jumping velocity is minimized for an optimum length. Figure S7B shows that while the substrate stiffness reduces, substrate frequency ratio increases monotonically with substrate length. Thus, an optimum combination of the two, obtained at an optimum length of ~ 15 mm, results in the minimum jumping velocity.

Figure S7A also shows comparison between CFD and the semi-empirical model in terms of jumping velocity ratio prediction. The semi-empirical model is able to successfully capture the trend of jumping velocity as a function of substrate length. Subsequently, we compare the two modeling approaches over a wide range of stiffness and frequency ratio values in Figure S7C. The two approaches are consistent for velocity ratio prediction over a wide range of parameter space. Thus, the semi-empirical model can be used to tune the substrate geometry and material to realize the minimization in jumping velocity.

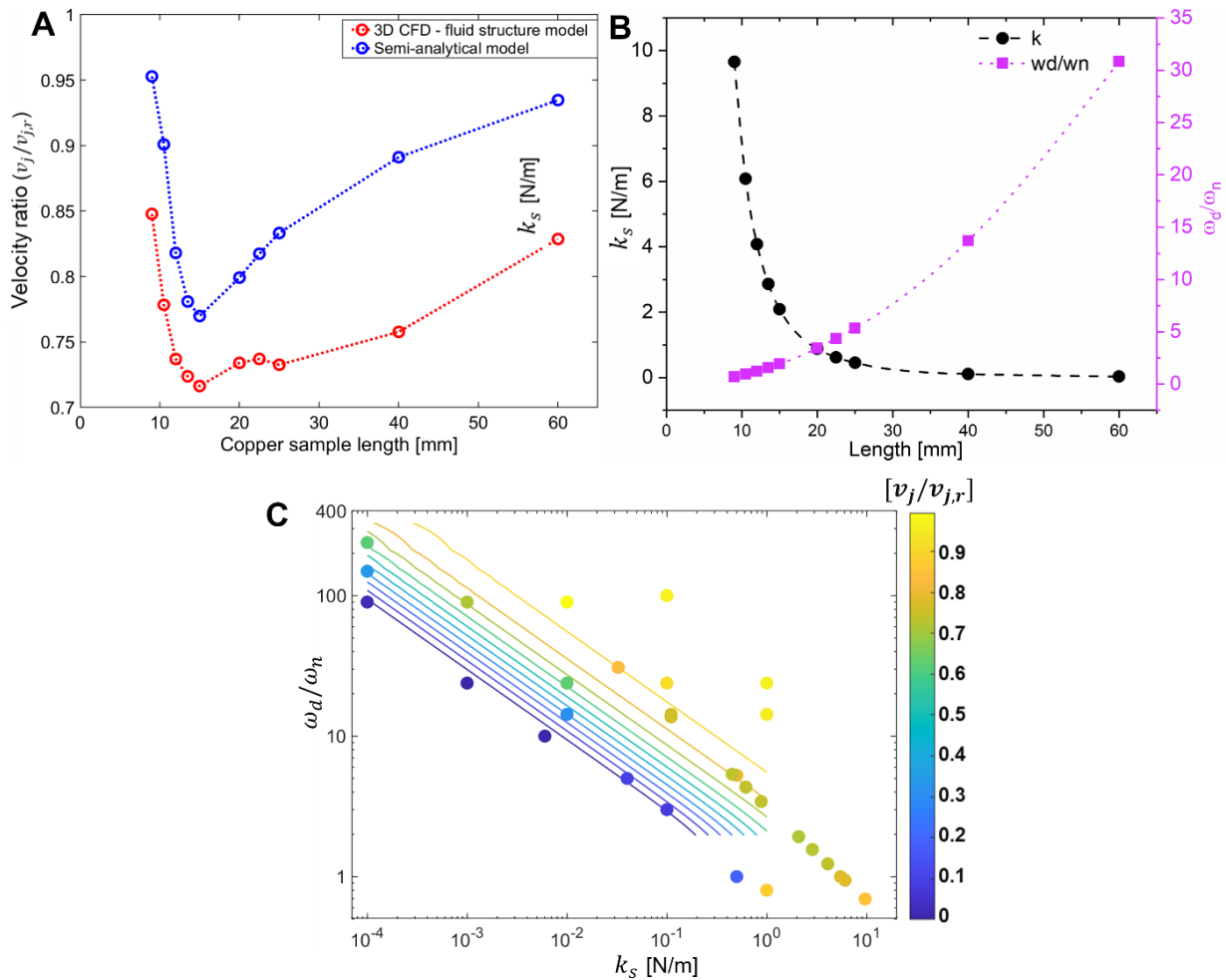

*Figure S7 A: Effect of varying the length of a thin copper sheet, 4 mm wide and 10 μm thick. The red dots are results of three-dimensional CFD simulations of drop coalescence on these flexible substrates. The blue dots are results of the semi-empirical model B: Substrate stiffness and frequency ratio variation with length C: Comparison of three-dimensional CFD model (dots) and semi-empirical model (curves) across a wide range of substrate stiffness and frequency ratios in terms of predictions for velocity ratio*

## S8. Effect of droplet size on coalescence dynamics

Figure S8 shows the force exerted by coalescing drops of various sizes on a rigid substrate. The magnitude of the force exerted on the substrate reduces with reduction in the size of the coalescing droplets. However, the similar nature of force for different sizes of the coalescing droplets is indicating that the coalescence dynamics are overall similar across the droplet sizes.

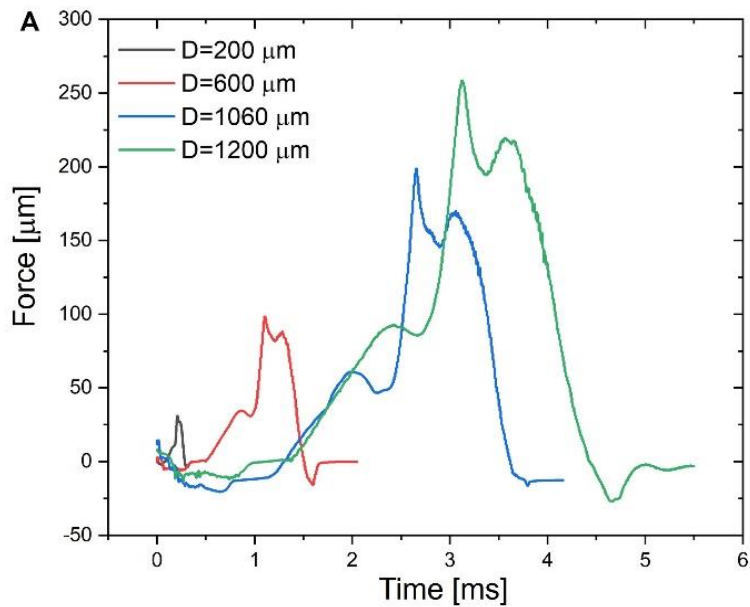

*Figure S8: Comparison of force exerted by coalescing droplets for various sizes of droplets showing a reduction in magnitude with the reduction in droplet diameter.*

**S9: Experimental details for fog harvesting**

In this setup, we have used copper foil backed by an aluminum plate as the rigid substrate and free-standing copper foil in vertical cantilever orientation as the flexible substrate, as shown in Figure S9. For the experiment with rigid substrate the substrate length is set as $L_2$ as marked in the figure. For the flexible substrate, the total length of the sample is set to $L_1+L_2$. The width of sample w remains the same for both rigid and flexible substrates. The distance between fog outlet and substrate is kept the same for a uniform fog density in both rigid and flexible substrate. The excess mass of substrate is removed in the case of flexible substrate to increase the natural frequency of substrate($\omega_n$), thus reducing the frequency ratio ($\omega_d/\omega_n$). For each experiment, the sample is suspended at the same distance $L_3$, above the collector plate. The collector plate is kept in a location as shown in Figure S9 and Figure 4c of the main paper. The location of the droplet falling on the substrate is captured through the CMOS camera (Thorlabs Inc.), as shown in the figure. All the experiments are carried out at a similar humidity in the room with a variation of 5-7%. The estimated stiffness ($k_s$) and natural frequency ($\omega_n$) value for the flexible substrate used in all the experiments is ~ $10^{-4} N/m$ and 52.2 rad/s respectively.

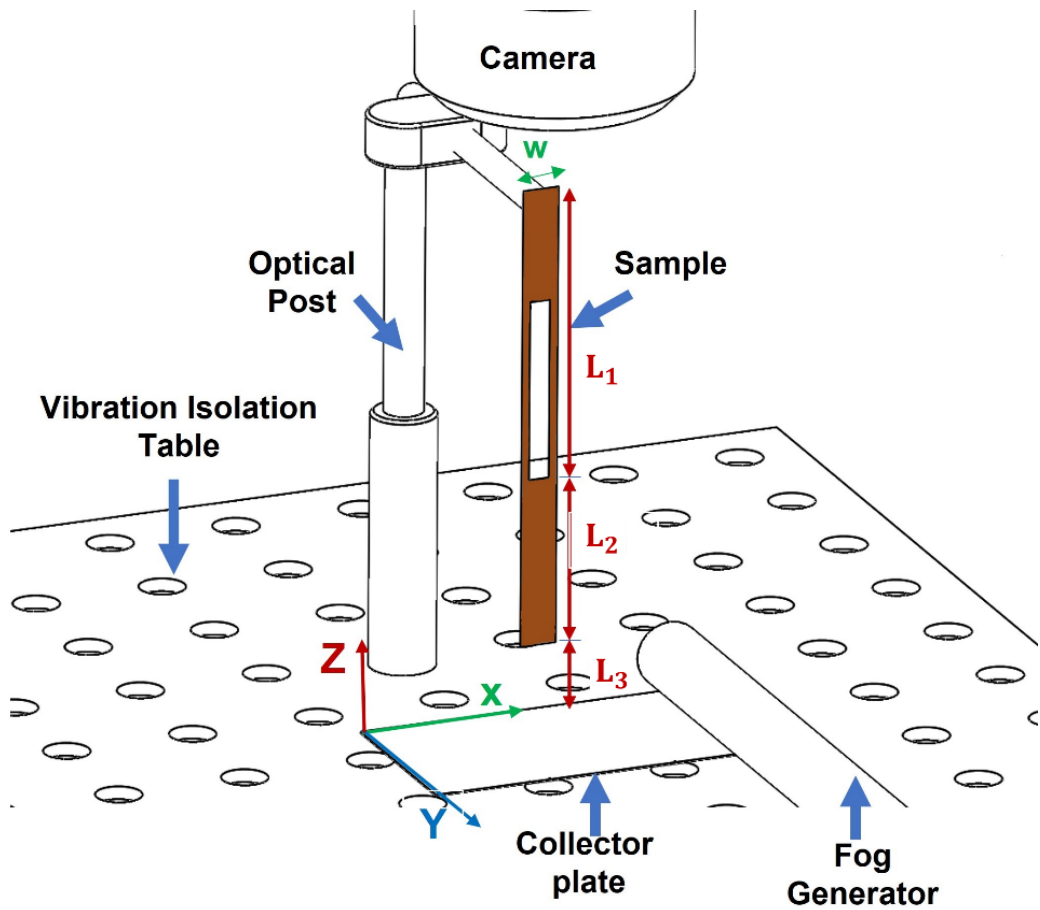

***Figure S9***: *Setup for water collection through fogging on superhydrophobic copper foil*

Figure S10 shows the comparison of the normalized area of droplets which are fallen on a 5mm strip starting from a distance $Y = 0$ for rigid and flexible cases. The normalized area is calculated by the ratio of total area of droplets on a particular 5 mm strip to the total area of all the droplets on the collector plate. For example, if we take a 5 mm strip between $Y = 10 \ and \ Y = 15$ in flexible substrate case, the total area of droplets on this strip is ~33% of total area of all the droplets deposited on the collector plate. We get Figure 4C in the main paper by cumulating the distribution shown in Figure S10.

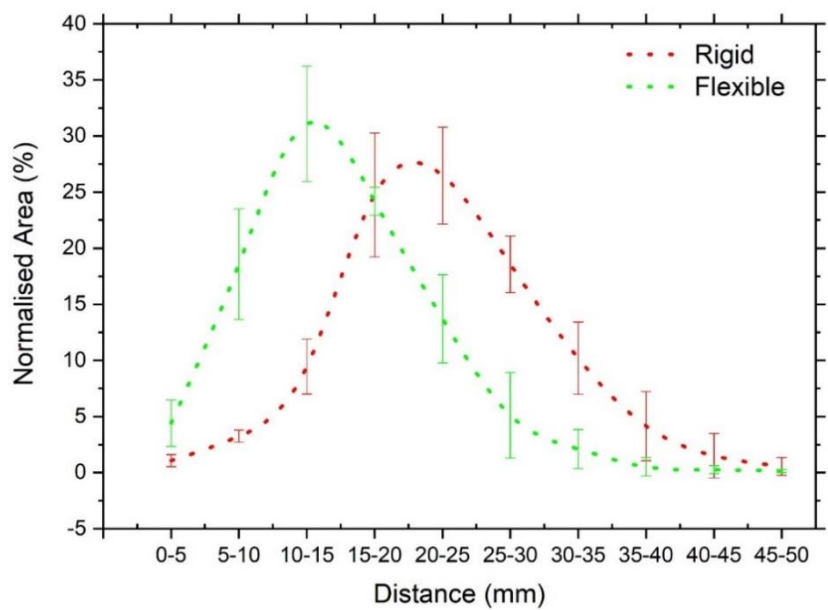

*Figure S10*: *Droplet distribution on the collector as a function of distance from the substrate for rigid (Red) and flexible (green) substrate (Y=0). The normalized area represents the percentage of total area of the droplet that has fallen on a 5mm strip at a particular distance (Shown in x-axis).*